\documentclass[preprint, 10pt, sort]{elsarticle}
\usepackage[utf8]{inputenc}
\usepackage[english]{babel}
\usepackage{textcomp}
\usepackage{bm}
\usepackage{lipsum,pdflscape}

\usepackage{graphicx}
\usepackage{subcaption}
\usepackage{amsmath,amssymb,mathtools,wasysym,array}
\DeclarePairedDelimiter{\ceil}{\lceil}{\rceil}
\usepackage[version=4]{mhchem}
\usepackage{setspace}
\usepackage{longtable,tabularx,multirow}
\newcommand\setrow[1]{\gdef\rowmac{#1}#1\ignorespaces}
\newcommand\clearrow{\global\let\rowmac\relax}
\clearrow
\usepackage[dvipsnames]{xcolor}
\usepackage{placeins,afterpage}
\usepackage{siunitx}

\usepackage{hyperref}
\makeatletter
\providecommand{\doi}[1]{%
  \begingroup
    \let\bibinfo\@secondoftwo
    \urlstyle{rm}%
    \href{http://dx.doi.org/#1}{%
      doi:\discretionary{}{}{}%
      \nolinkurl{#1}%
    }%
  \endgroup
}
\makeatother

\usepackage{url}

\usepackage[capitalise,noabbrev]{cleveref}
\crefrangelabelformat{equation}{(#3#1#4-#5#2#6)}

\usepackage[acronym]{glossaries}
\setacronymstyle{long-short}
\newacronym{leo}{LEO}{Low Earth Orbit}
\newacronym{pnp}{PNP}{Probability of No-Penetration}
\newacronym{lmf}{LMF}{Liquid Mass Fraction}
\newacronym{ble}{BLE}{Ballistic Limit Equation}
\newacronym{drama}{DRAMA}{Debris Risk Assessment and Mitigation Analysis}
\newacronym{nsga2}{NSGAII}{Non-dominated Sorting Genetic Algorithm 2}
\newacronym{dod}{DOD}{depth-of-discharge}
\newacronym{deap}{DEAP}{Distributed Evolutionary Algorithms in Python}
\newacronym{srl}{SRL}{Schafer-Ryan-Lambert}
\newacronym{master}{MASTER}{Meteoroid and Space Debris Terrestrial Environment Reference}
\newacronym{cfrp}{CFRP}{Carbon Fibre Reinforced Plastics}
\newacronym{hcsp}{HC-SP}{Honeycomb Sandwich Panel}
\newacronym{fp}{FP}{Flat Plate}

\begin{document}

\begin{frontmatter}
\title{Constrained optimisation of preliminary spacecraft configurations under the design-for-demise paradigm \fnref{fn3}}
\author{Mirko Trisolini\corref{cor1}\fnref{fn1}}
\ead{mirko.trisolini@polimi.it}
\address{University of Southampton, University Road, SO17 1BJ, Southampton, United Kingdom}
\author{Hugh G. Lewis}
\ead{H.G.Lewis@soton.ac.uk}
\address{University of Southampton, University Road, SO17 1BJ, Southampton, United Kingdom}
\author{Camilla Colombo}
\ead{camilla.colombo@polimi.it}
\address{Politecnico di Milano, Via La Masa 34, 20156, Milano, Italy}
\cortext[cor1]{Corresponding author}
\fntext[fn1]{Present address: Politecnico di Milano, Via La Masa 34, 20156, Milano, Italy}

\fntext[fn3]{Accepted for publication in the Journal of Space Safety Engineering. doi: 10.1016/j.jsse.2021.01.005}


\begin{abstract}
	In the past few years, the interest towards the implementation of design-for-demise measures has increased steadily. The majority of mid-sized satellites currently launched and already in orbit fail to comply with the casualty risk threshold of \text{$10^{-4}$}. Therefore, satellites manufacturers and mission operators need to perform a disposal through a controlled re-entry, which has a higher cost and increased complexity. Through the design-for-demise paradigm, this additional cost and complexity can be removed as the spacecraft is directly compliant with the casualty risk regulations. However, building a spacecraft such that most of its parts will demise may lead to designs that are more vulnerable to space debris impacts, thus compromising the reliability of the mission. In fact, the requirements connected to the demisability and the survivability are in general competing. Given this competing nature, trade-off solutions can be found, which favour the implementation of design-for-demise measures while still maintaining the spacecraft resilient to space debris impacts. A multi-objective optimisation framework has been developed by the authors in previous works. The framework's objective is to find preliminary design solutions considering the competing nature of the demisability and the survivability of a spacecraft since the early stages of the mission design. Multi-objective optimisation is used to explore the large search space of the possible configurations in order to find a range of optimised trade-off solutions that can be used for the future phases of the mission design process. In this way, a more integrated design can be achieved. The present work focuses on the improvement of the multi-objective optimisation framework by including constraints. The constraints can be applied to the configuration of the spacecraft, with limitations on the location of the internal components, or to specific components, based on their feasibility and design. The evaluation of the demisability and survivability is carried out with two dedicated models, and the fitness of the solution is assessed through two indices summarising the level of demisability and survivability. 
\end{abstract}


\end{frontmatter}


\section{Introduction}
With the increased attention towards a more sustainable use of space that has been developing in recent years, all the major space-faring nations and international committees have proposed a series of debris mitigation guidelines to ensure the future exploitation of the space environment \cite{OConnor2008,Schafer2005}. One of the key aspects of these guidelines is the removal of dismissed satellites from the protected regions as soon as possible, and in a maximum of 25 years \cite{EuropeanSpaceAgency2008,NASA2012_process}. If we focus on the \gls{leo} environment, the disposal of dismissed satellites is achieved by de-orbiting them within 25 years from the end of their operational life. However, the consequence of this process is the re-entry of satellites through Earth's atmosphere. This can pose a risk for people and properties on the ground as part of the satellite components and structures can survive the re-entry process. According to standard regulations, the casualty risk for people on the ground cannot exceed the value of \text{$10^{-4}$} \cite{NASA2012_process,EuropeanSpaceAgency2008}. One way in which this risk can be mitigated is through a design philosophy, which is referred to as \emph{design-for-demise}. Within this paradigm, the satellite is designed in a way that most of its parts will be demised in the atmosphere thus limiting the casualty risk on ground. The implementation of design-for-demise strategies does not only consider the safety of people on ground, but also favours the selection of uncontrolled disposal strategies, which are simpler and cheaper alternatives with respect to controlled re-entries \cite{Waswa2012,Waswa2013}. As mentioned, designing for demise influences the design of a spacecraft and its components as it directly acts on the choices of their material, shape, dimensions, position to obtain a more demisable solution. However, a spacecraft designed to demise still has to survive the space environment for many years, which is populate by a large number of space debris that can impact the satellite and damage its structures and components \cite{Christiansen2009,Putzar2006,Grassi2014}. Consequently, the design of a spacecraft must consider such requirements. In addition, these requirements should be considered from the early stages of the mission design. Considering them at a later stage may cause inadequate integration of these design solutions, leading to a delayed deployment of the mission and to an increased cost of the project \cite{Waswa2013}. Demisable designs will tend to favour lighter materials, thinner structures, and more exposed components, whereas survivability oriented designs will favour denser materials, thicker structures, and more protected components. It is evident that these requirements are competing \cite{Trisolini2018_Acta}; consequently, it was decided to develop a multi-objective optimisation framework that would consider both this aspects and help finding trade-off preliminary satellite configurations. These solutions can then be used as a starting point for a more detailed design. As the problem is non-linear and involves the combination of continuous and discrete variables, classical derivative based approaches are unsuited and a genetic algorithm was selected instead. Moreover, the framework relies on two models \cite{Trisolini2018_AESCTE,THC2020_Survivability} to assess the demisability and the survivability as a function of several design parameters, and on two indices, which are used to evaluate the level of demisability and survivability and serve as fitness functions in the optimisation process.
\bigbreak
The present work builds on the results presented in \cite{Trisolini2018_AESCTE}, where the optimisation framework was firstly introduced and an unconstrained optimisation was performed on a relevant test case. In the present contribution, the framework has been further developed, in particular with the introduction of components and mission specific constraints. The improved multi-objective optimisation has then been applied to the test case already discussed in \cite{Trisolini2018_AESCTE} and the results show a clear difference now that the constraints on the components have been introduced. A second test case is also presented, which shows the application of the framework to a more complex configuration. The modelled satellite contains tanks, reaction wheels, batteries, payloads, which are all considered together with their respective constraints. Such a test case demonstrates how the framework behaves with a more complex problem and with several optimisation variables of mixed type.


\section{Demisability model}
\label{sec:demisability}
The following section describes an \emph{enhanced} object-oriented destructive re-entry code \cite{NASA2009_ORSAT,NASA2015_DAS,Lips2005a} developed at the University of Southampton used for the assessment of the demisability of re-entering spacecraft. The enhanced feature of the model resides in the additional capabilities with respect to standard object oriented codes, such as the possibility of modelling the early detachment of the external panels. In addition, the Phoenix tool has been developed to ensure compatibility with the survivability analysis tool with which is used in combination in the multi-objective optimisation framework. The main features of these types of codes are:

\begin{itemize}
	\item Hierarchical spacecraft definition;
	\item Schematisation of components and structures with primitive shapes (cube, cylinder, sphere, and flat plate);
	\item Use of engineering correlations for aerodynamics and aerothermodynamics interactions.
\end{itemize} 

The first hierarchical level is the main spacecraft structure (the \emph{parent object}). Here the overall spacecraft mass and dimensions are specified. In addition, the solar panels can be defined and schematised as flat plates. The re-entry simulation of the solar panels is not performed and they are assumed to demise; however, their area is taken into account in the computation of the aerodynamic cross-section of the satellite until they detached at a user specified altitude \cite{Gelhaus2014}. The second level defines the external panels of the main structure. This allows the definition of different characteristics for each external panel, such as the material, the thickness, and the type of panel itself (e.g. single wall, double wall, and honeycomb sandwich panel). This is one of the features that separates the developed model from traditional object-oriented codes. The third level defines the main internal components and subsystems such as tanks and reaction wheels assemblies. When internal components are defined, it is possible to attach them to the external panels defined in the second hierarchical level. This allows considering an additional demise mechanism: taking into account the early-release (before the main break-up) of the attached components. An additional level can also be used for the definition of sub-components such as battery cells. An example with a small configuration is shown in \cref{tab:config}. For readability reasons not all the parameters required for the definition of a component are presented in the table. 

\begin{table}[hbt!]
\caption{\label{tab:config} Example of spacecraft configuration required by the survivability model.}
\centering
\begin{tabular}{>{\rowmac}l>{\rowmac}l>{\rowmac}c>{\rowmac}c>{\rowmac}c>{\rowmac}c>{\rowmac}c>{\rowmac}c>{\rowmac}c>{\rowmac}c>{\rowmac}c<{\clearrow}}
\hline
\setrow{\bfseries}ID & Name & $\mathbf{ID_{p}}$ & Shape & \boldmath{$m$} & \boldmath{$l$} & \boldmath{$r$} & \boldmath{$w$} & \boldmath{$h$} & \boldmath{$n$} & \\
 & & & & (\si{\kilo\gram}) & (\si{\meter}) & (\si{\meter}) & (\si{\meter}) & (\si{\meter}) & & \\ \hline
0 & Spacecraft & n/a & Box & 2000 & 3.5 & n/a & 1.5 & 1.5 & 1 & \\
1 & Tank & 0 & Sphere & 15 & n/a & 0.55 & n/a & n/a & 1 & \\
2 & BattBox & 0 & Box & 5 & 0.6 & n/a & 0.5 & 0.4 & 1 & \\
3 & BattCell & 2 & Box & 1 & 0.1 & n/a & 0.05 & 0.04 & 20 & \\
\hline
\end{tabular}
\end{table}

When the re-entry simulation is performed, the first part only takes into account the parent spacecraft, until the main break-up altitude is reached. In this phase, only the parent structure can interact with the external heat flux. It is assumed that the internal components do not experience any heat load during this phase. The break-up altitude is user-defined. During this first phase, also the occurrence of specific events is considered that is the detachment of the solar panels and of the external panels of the main structure. The solar panels are considered to detach at a predefined user altitude; standard values range in the interval 90-95 km \cite{Gelhaus2014,Gelhaus2013,Martin2005}. When it is reached, the solar panels are simply removed from the simulation, with the consequent change in the aerodynamics of the main body. The detachment of the external panels of the structure is instead triggered by a temperature condition. Once the panels reach the melting temperature, they are assumed to detach \cite{Koppenwallner2005}. This is a conservative approach as the panels may detach earlier because of the melting of the bolts or of the adhesive, which are used to mount them. If an internal component is attached to the panel, it is released when the panel detaches. The first part of the simulation ends with the main spacecraft reaching the break-up altitude. At this point, both the internal objects and the remaining external panels are detached from the main structure and released. The trajectory is then simulated separately for each component. Finally, the final mass, cross-section, landing location and impact energy of each surviving object, and the demise altitude of each demised object is stored. A more complete description of the demisability model and its comparison with state-of-the-art destructive re-entry codes, can be found in \cite{Trisolini2015_Jerusalem,Trisolini2016_JSSE,Trisolini2017_Adelaide,Trisolini2018_Acta,Trisolini2018_AESCTE}.


\section{Survivability model}
\label{sec:survivability}
The following section describes a spacecraft survivability model, which analyses the resistance against the impact from untraceable space debris. As both the survivability and the demisability model are combined in a multi-objective optimisation framework, they need to share the same input configuration. Consequently, the definition of the configuration coincides with the one of the demisability model (\cref{sec:demisability}), with the added requirements that the position and orientation of the internal components have to be provided. This model considers the overall configuration (external structure and internal components) as made of a collection of panels, with the relative physical and geometrical properties associated to them. The main body of the satellite can be defined using a single material and a uniform thickness or different materials, thickness, and shielding types can be defined for each face. Three types of shielding can be used: single wall shields, Whipple shields, and honeycomb sandwich panels \cite{Christiansen2009}. The components can be positioned free inside the main structure, or be attached to the external panels. As the model performs preliminary vulnerability analyses, it has been decided to specify the orientation only along the main axes of the satellite. Alongside the geometrical representation of the satellite, the mission orbit and the characteristics of the space environment need to be known. Using the operational orbit, the debris environment surrounding the spacecraft is schematised through \emph{vector flux elements} \cite{Trisolini2018_Acta}, which carry the information on the debris flux, particle diameter, impact velocity, impact azimuth, and impact elevation for different sectors of the space surrounding the satellite. The vector flux elements are then used in conjunction with the geometrical characteristics of the spacecraft and the \glspl{ble} \cite{Schafer2008,Ryan2010,Ryan2011} to compute the penetration probability on each panel of the structure. The penetration probability is computed assuming that the impact of each particle is statistically independent, so that a Poisson distribution can be adopted for the computation \cite{Bunte2017,Gade2013,Welty2013,Grassi2014}. 

The vulnerability assessment of the external structure is different from the one for the internal components. For the former, the direct impact of the debris particles is considered, whereas for latter, the debris clouds that develop inside the spacecraft after impacting the external structure is considered \cite{Christiansen2009,Grassi2014,Putzar2006,Welty2013}. In fact, these clouds can hit and damage internal components. To model this phenomenon, the \gls{srl} \gls{ble} \cite{Grassi2014,Welty2013} has been used, which takes into account impacts on multi-walled structures. The procedure is also based on the concept of vulnerable zones \cite{Trisolini2018_AESCTE,THC2020_Survivability,Putzar2006}. They consist of an adjusted projection of an inner component onto the outer spacecraft structure. This area represents the portion of the external structure that, if impacted by a particle, could lead to the impact on the inner component to which the relevant vulnerable zone is associated. In addition, the model is able to consider the effect of the mutual shielding between internal components. In fact, the effect of the resulting internal propagation of the ejecta can be significantly influenced by the presence of interposing components between the impacted panel and the target component. A more comprehensive description of the survivability model and its features can be found in \cite{Trisolini2015_Jerusalem,Trisolini2016_JSSE,Trisolini2017_Adelaide,Trisolini2018_Acta,Trisolini2018_AESCTE,THC2020_Survivability}.


\section{Multi-objective optimisation framework}
\label{sec:opt_framework}
A multi-objective optimisation framework has been developed to identify trade-off solutions considering the demisability and survivability of satellite configurations \cite{Trisolini2018_AESCTE}. The objective is to obtain options for more integrated configurations since the early stages of the mission design. In fact, a spacecraft designed to demise will have thinner and more exposed structures, made of lighter materials; however, it still has to survive the debris environment for many years. To do so, the optimisation framework uses the demisability and survivability models described in \cref{sec:demisability} and \cref{sec:survivability}. The results from both these analyses are in general competing \cite{Trisolini2018_Acta}; consequently, optimized solutions represent a trade-off between demisable and resilient solutions. When a satellite configuration with several components is analysed the task become complex and many parameters are considered. From the geometry and size of the components, to their material and positioning inside the spacecraft. In addition, the components have to satisfy specific constraints, which may be related to the mission design or specific to the component itself. The developed multi-objective optimisation framework considers all these aspects, allowing the evaluation of many possible preliminary configurations and the comparison amongst them, providing a wide range of viable solutions in the initial phases of the mission design.

A broad spectrum of optimisation strategies can be used for this task; however, for the purpose of this work and for the characteristics of the problem in exam, genetic algorithms have been selected. Their selection is based on their strong heritage in solving non-linear optimisation problems with large search spaces and with mixed types of variables (integer and real). In addition, we have already successfully used and tested them in a previous work \cite{Trisolini2018_AESCTE}. The \gls{deap} \cite{Fortin2012_deap} Python framework was used for the development of the optimisation framework. The adopted selection strategy is the \gls{nsga2} \cite{Deb2002_NSGA}, the selected crossover mechanism  is the Simulated Binary Bounded \cite{Deb1995_crossover} and the mutation mechanism is the Polynomial Bounded \cite{Deb2014_mutation}. The following sections present the main characteristics of the framework.

\subsection{Generation of the individuals of the population}
\label{subsec:individuals}
During the optimisation, each satellite configuration is represented by an individual, which contains the \emph{genetic material} defining a configuration. A real valued representation for the individuals has been selected \cite{Deb2001}. With this representation, each gene directly represents the variable to be optimized. The variables can be both integer and floating-point numbers. For continues variables such as the thickness, the floating representation is used, whereas for discrete variables such as the material, the integer representation is adopted. \cref{tab:opt_variables} summarises all the possible variables that can be optimized in the developed framework, together with their representation in the individual.

\begin{table}[hbt!]
\caption{\label{tab:opt_variables} Summary of possible optimisation variables.}
\centering
\begin{tabular}{lcc}
\hline
\textbf{Variable} & \textbf{Variable type} & \\ \hline
Component quantity & Integer & \\
Component material & Integer & \\
Component shape & Integer & \\
Component thickness & Real & \\
Component parent & Integer & \\
Component position & Real & \\
Component size & Real & \\
Component orientation & Integer & \\
External panel material	& Integer & \\
External panel shielding & Integer & \\
External panel thickness & Real & \\
\hline
\end{tabular}
\end{table}

The variable \emph{quantity} identifies how many instances of a component will be generated. The \emph{component parent} is used to specify the hierarchical relation between a component and another one through their IDs. This is a connection between the two components, which can be a containment if the ID refers to another internal component or attachment if the ID identifies one of the external panels. The \emph{orientation} is represented with an integer as only predefined orientations along the main axis of the spacecraft are be considered. The \emph{component catalogue} variable can be to provide catalogued options for certain type of components as, for example, battery cells. The \emph{external panel-shielding} takes into account the possibility to have external panels made of a single panel sheet, Whipple shields, or \gls{hcsp}. For each component to be optimized in the configuration, the user can specify which variables will be optimized. For each optimisation variable is also necessary to specify the lower and upper bounds from which the optimizer can sample during the generation of the population. For example, the \emph{material} can be selected from an available database (\cref{tab:mat_database}), where to each material is associated an identifier. The \emph{size}, \emph{thickness}, and \emph{position} can be component and mission dependent and can be set separately for each component. 

\subsection{Constraints handling and components requirements}
\label{subsec:constraints}
When designing spacecraft configurations, it is necessary to consider specific constraints to ensure their feasibility. For example, tanks must be strong enough to sustain a specific internal pressure. These types of constraints need to be addressed inside the optimisation as a randomly generated individual may not be feasible. When the feasibility is checked, two possibilities are available: the individual is either rejected or repaired through a specific repairing function. For each component, is also possible to specify limits on several of the available optimisation variables, such as the material, the shape, or the parent object of the component. For example, the attachment of a component can be constrained only to specific external panels, or to specific regions inside the satellite. This last aspect, for example, can be useful whenever components must be positioned on the Earth-facing side to make observations. In the current implementation of the framework, the constraints handling procedure is implemented for tanks, reaction wheels, and batteries; however, user-defined constraints can always be added to each component.

\subsubsection{Tank assemblies constraints}
\label{subsubsec:tank_constraints}
For the feasibility of a tank assembly it is necessary to consider the the propellant mass, the storage pressure, the number of tanks, their shape, thickness, and material. These variables are critical for the realistic definition of a tank assembly. Starting from the amount of propellant required by the mission, the tank volume $V_t$ is computed as

\begin{equation} \label{eq:fuel_volume}
V_t = K_1 \cdot \frac{m_f}{\rho_f}
\end{equation}

where $\rho_f$ is the density of the fuel, $m_f$ is the mass of the fuel for the entire mission, and $K_1$ is a factor that takes into account the additional volume needed for the pressurant gas (the filling factor) \cite{AirbusSafranLaunchersGmbH2003}. Given the propellant volume, the shape, and the number of tanks we can then compute the internal radius of the vessel. For spherical tanks

\begin{equation} \label{eq:rint_sphere}
r_{t,i}^{s} = \sqrt[3]{\frac{3 \cdot V_t}{4\pi \cdot n_t}}
\end{equation}

where $r_{t,i}^{s}$ is the internal radius of the spherical tank and $n_t$ is the number of tanks in the configuration. For cylindrical tanks, the equivalent expression is

\begin{equation} \label{eq:rint_cyl}
r_{t,i}^{c} = \sqrt[3]{\frac{V_t}{2\pi \cdot A\!R \cdot n_t}}
\end{equation}

where, $r_{t,i}^{c}$ is the internal radius and $A\!R$ is the aspect ratio of the cylinder (ratio between the length and the diameter). Finally, we compute the ultimate strength acting on the walls of the tank. For spherical tanks, it can be expressed as follows:

\begin{equation} \label{eq:sigmau_cyl}
\sigma_{u,w}^{s} = \frac{S\!F \cdot p \cdot r_{t,i}^{s}}{2 \cdot t_s}
\end{equation}

Whereas for cylindrical tanks the equivalent expression is

\begin{equation} \label{eq:sigmau_sphere}
\sigma_{u,w}^{c} = \frac{S\!F \cdot p \cdot r_{t,i}^{c}}{t_s}
\end{equation}

where $t_s$ is the thickness of the tank (considered uniform), \textit{p} is the operating pressure, and \textit{SF} a safety factor. A solution is feasible if the ultimate strength of the material ($\sigma_{u,m}$) is grater than the one acting on the tank walls as follows:

\begin{equation} \label{eq:sigma_comparison}
\sigma_{u,w} < \sigma_{u,m} 
\end{equation}

The implementation of the constraint inside the multi-objective optimisation algorithm is in the form of a death penalty \cite{Coello2002}, where infeasible solutions are directly discarded.

When initialising the optimisation problem some data must be provided for the feasibility check, specifically, the propellant mass, the storage pressure, the fuel density, the safety factor, and the filling factor, which are all dependent on the specifics of the mission scenario. In addition, the aspect ratio of the tank can be provided or optimized. However, the sensitivity of the demisability and survivability indices to the aspect ratio is low with respect to the other parameters considered \cite{Trisolini2018_Acta}.

\subsubsection{Reaction wheels constraints}
\label{subsubsec:rw_constraints}
For the reaction wheels, two separate checks must be performed. First, the amount of angular momentum provided by the reaction wheel must be greater than the minimum required by the mission, and second, the structural integrity must be assured. The minimum required radius for the reaction wheel to satisfy the angular momentum requirements can be computed as follows \cite{Wertz1999}

\begin{equation} \label{eq:rmin_wheel}
r_{rw,min} = \bigg( \frac{H_{d}}{\pi \cdot \rho_m \cdot \omega_{\rm max} \cdot A\!R} \bigg)^{1/5}
\end{equation}

where $H_{d}$ is the design angular momentum of the wheel, $\rho_m$ is the material density, $\omega_{\rm max}$ is the maximum rotation speed of the wheel, and $A\!R$ is the aspect ratio of the wheel, which is always assumed to by cylindrical in shape. The structural integrity can then be checked as follows \cite{Wertz1999}

\begin{equation} \label{eq:sigmay_wheel}
\sigma_y \leq S\!F \cdot (3+\nu) \cdot \rho_m \cdot \omega_{max}^2 \cdot r_{rw}^2
\end{equation}

where $\sigma_y$ is the yield strength of the material, $S\!F$ is a safety factor, $r$ is the radius of the wheel, and $\nu$ is the Poisson ratio of the material.
If the radius of the reaction wheel does not satisfy the requirements, it is repaired and the radius is set equal to the minimum possible value, as long as such value is inside the initially specified boundaries. On the other end, if the structural integrity test fails, the solution is discarded.

\subsubsection{Batteries requirements}
\label{subsubsec:batt_requirements}
For battery cells, the optimisation procedure considers the possibility of selecting them from a catalogue. This is the most common way to proceed during an actual mission design, where battery cells are selected based on their characteristics, such as \gls{dod}, capacity, voltage, mass, and dimension, and their quantity and disposition is derived from the mission requirements, i.e. the power required, the eclipse time, and the required voltage. Once they are selected, the optimiser checks the requirements of the missions against the selected battery cell and performs a preliminary design to get the number of cells needed for the mission. Following the simplified procedure outlined in \cite{Wertz1999}, first the required total battery capacity ($C$) is computed as

\begin{equation} \label{eq:capacity}
C = \frac{W_e \cdot T_e}{\eta \cdot N_b \cdot D\!O\!D}
\end{equation}

where $W_e$ is the power required by the spacecraft during eclipse in Watts, $T_e$ is the time spent in eclipse in hours, $\eta$ is the transmission efficiency (assumed equal to 0.9 \cite{Wertz1999}), and $N_b$ is the number of batteries (not battery cells). Given the battery capacity and its specific energy density, the mass of the battery can be obtained as follows

\begin{equation} \label{batt_mass}
m_b = \frac{C}{E_d} \cdot N_b
\end{equation}

where $E_d$ is the specific energy density of the battery in Wh/kg. The number of cells required for the battery is then estimated as follows

\begin{equation} \label{batt_cells}
n_c = \ceil[\Big]{\frac{m_b}{m_c}}
\end{equation}

where $m_c$ is the mass of the battery cell, which can be obtained from a catalogue.

\subsubsection{Position errors handling}
\label{subsubsec:pos_constraints}
\begin{figure}[htb!]
\centering
\includegraphics[width=0.65\textwidth, keepaspectratio]{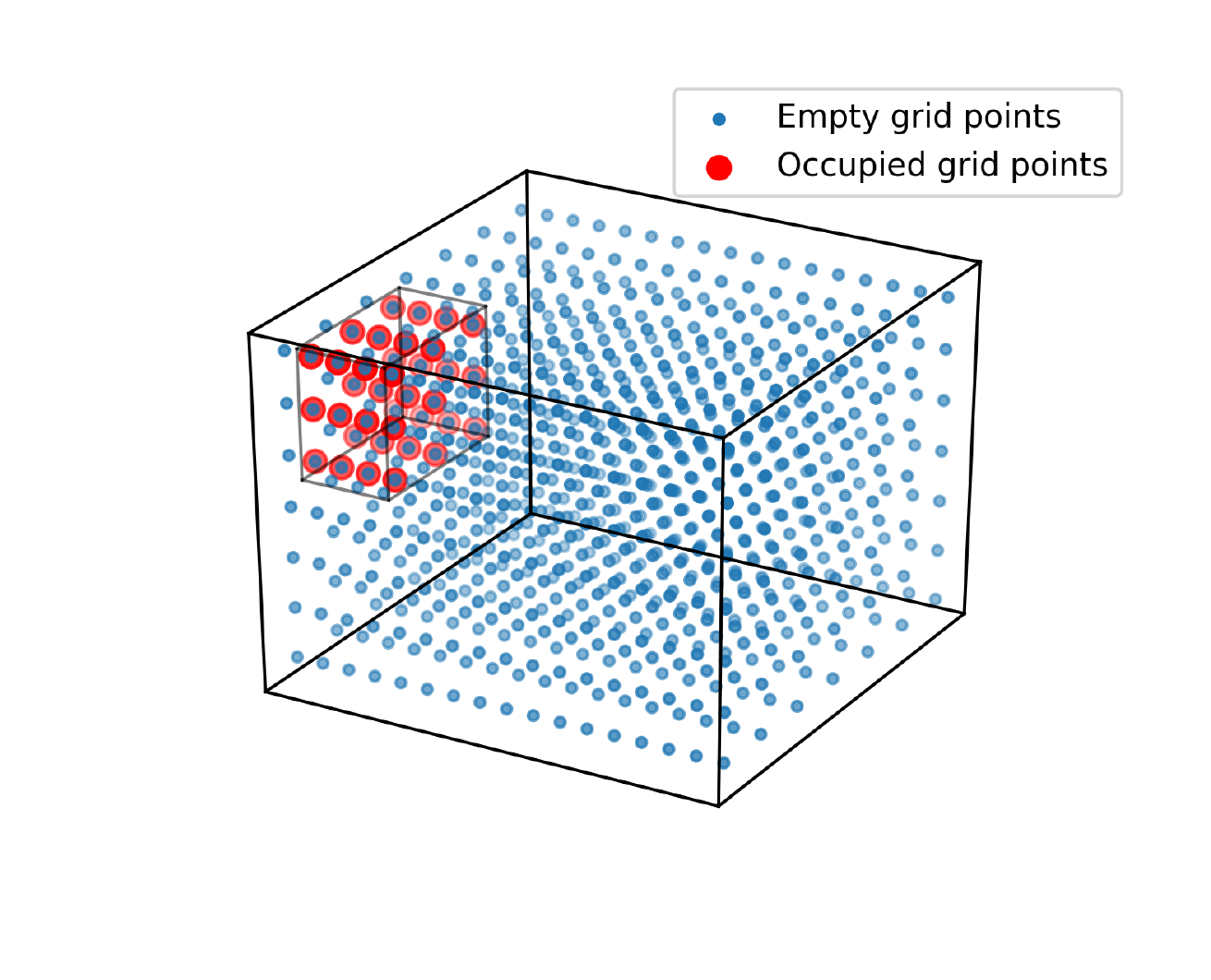}
	\caption{Position grid representation with occupied points.}
	\label{fig:position_grid}
\end{figure}

When the position of the components can be optimized, the random generation of the individual can result in a configuration with overlapping components. As this is a rather common event, it was decided not to discard these solutions, but to repair them, moving the components until they no longer intersect each other. The procedure relies on the generation of a 3D grid, which subdivides the interior part of the parent spacecraft. When the configuration is generated, all the grid points are flagged with a 0, meaning that all points are available. Every time a component is added to the grid, the occupied grid points switch to 1, meaning that they are no longer available (\cref{fig:position_grid}) to be occupied by other components. Following this simple procedure, the configuration of the satellite is defined in an incremental fashion.

When the components overlap, the repairing procedure must be carried out. The procedure consists of the following steps:
\begin{itemize}
\item
Identification of the intersecting components;
\item
Selection of one of the components to be repaired;
\item
Addition to the grid of all the components not involved in the intersection;
\item
Sorting of all the available grid points with respect to the distance from the original component position;
\item
Positioning of the component in the closest available grid points;
\end{itemize}

The procedure is then repeated until all intersecting components are repaired and the configuration is feasible. In case the repairing algorithm is not successful, the solution is discarded. The available grid points are sorted with respect to the distance from the original position of the component to maintain the highest similarity with the randomly generated individual.

\subsection{Fitness functions definition}
\label{subsec:fitness}
To evaluate the level of demisability of a configuration, the \gls{lmf} \cite{Trisolini2018_AESCTE} index is introduced. The \gls{lmf} index represents the ratio between the mass that melts during the re-entry and the initial mass of the spacecraft. The \gls{lmf} index can be expressed as

\begin{equation} \label{eq:lmf}
L\!M\!F = 1 - \frac{\sum_{j=1}^{N} m_{fin,j}}{\sum_{j=1}^{N} m_{in,j}}
\end{equation}

where $m_{fin,j}$ and $m_{in}$, \textit{j} are the final and initial mass of the \textit{j}-th component respectively, and \textit{N} is the total number of components. Only the internal components are considered in the computation of the index. The external structure is assumed to demise.

To evaluate the level of survivability, the \gls{pnp} \cite{Trisolini2018_AESCTE} is used. The probability of no-penetration represents the chance that a specific spacecraft configuration is not penetrated by space debris during its mission lifetime. In this case, the penetration of a particle is assumed to produce enough damage to the components to seriously damage them so that the \gls{pnp} can be considered a sufficient parameter to evaluate the survivability of a satellite. The overall probability of no-penetration of a spacecraft configuration is given by

\begin{equation} \label{eq:pnp}
P\!N\!P = 1 - \sum_{j=1}^{N} P_{p,j}
\end{equation}

where $P_{p,j}$ is the penetration probability on the \textit{j}-th component. As for the demisability index, also in this case only the internal components are considered. In case sub-components are presents, i.e. component contained inside internal components such as battery cells inside a battery box, only the \gls{pnp} on the container is considered.


\section{Test case: Tank assembly}
\label{sec:test1}
This test case aims at comparing the results of the optimisation with and without the application of the constraints. For this purpose, we perform the constrained optimisation on the same test case presented in \cite{Trisolini2018_AESCTE}, where the constraints where not implemented yet in the framework. Despite the mission scenario and spacecraft configuration are the same as in \cite{Trisolini2018_AESCTE}, the optimisation simulations have been re-run also for the unconstrained case, because new features have been introduced in the survivability model \cite{THC2020_Survivability}. In the following, we give an overview of the characteristics of the test case; for a more detailed description the reader can refer to \cite{Trisolini2018_AESCTE}.
\bigbreak
The test case features the optimisation of tank assemblies. They have been selected as they are critical in term of their demisability as they usually survive re-entry, and they also need protection from debris impacts as they are pressurised components critical to the mission success. As the optimisation is focused on a subsystem of the spacecraft, it is necessary to define the other characteristics of the configuration such as the external structure. For this simplified test case a cubic external structure is selected with a side length proportional to the mass of the spacecraft, $ m_s $, as follows \cite{Wertz1999}.

\begin{equation} \label{eq:side_length}
	L = \sqrt[3]{\frac{m_s}{\overline{\rho}_s}},
\end{equation}

where $ \overline{\rho}_s $ is the average density of the spacecraft, assumed equal to 100 \si{\kilo\gram/\meter\cubic} \cite{Wertz1999}. In addition to the size and mass of the spacecraft, the thickness and material of the external wall must be defined. For this test case, and a 3 mm Al-6061-T6 single wall as been selected.
\bigbreak
Alongside the external configuration, the mission scenario must be specified that is the operational orbit for the debris impact analysis and the re-entry initial conditions for the demisability assessment. The operational orbit is a sun-synchronous orbit at an altitude of 802 \si{\kilo\meter} and an inclination of 98.6\si{\degree}. For the re-entry initial conditions, an altitude of 120 \si{\kilo\meter}, flight path angle of 0\si{\degree}, velocity of  7.8 \si{\kilo\meter \per \second}, longitude of 0\si{\degree}, latitude of 0\si{\degree}, and heading angle of -8\si{\degree} have been selected.

\subsection{optimisation set-up}
\label{subsec:opt_var1}
The setup of the optimisation follows the one presented in \cite{Trisolini2018_AESCTE}, where the optimisation variables are the tank shape, thickness, material, and number of vessels. The size of the tanks (i.e. the radius) is instead related to the total volume of propellant needed by the mission and the number of tanks to have a realistic mission scenario \cite{Trisolini2017_Adelaide, Trisolini2018_AESCTE}. \cref{tab:opt_var_tank} summarises the variables taken into account and their bounds/options.

\begin{table}[!hbt]
\caption{\label{tab:opt_var_tank} optimisation variables for the tank assembly test case.}
\centering
\begin{tabular}{lcc}
\hline
\textbf{Variable} & \textbf{Bounds / Options} & \\ \hline
Material & Al-6061-T6, AISI-316, Ti-6Al4V & \\
Thickness & 0.0005 \si{\milli\meter} – 0.005 \si{\milli\meter} & \\
Shape & Sphere, Cylinder & \\
No. of vessels & 1 - 6 & \\
\hline
\end{tabular}
\end{table}

\begin{figure}[!htb]
 \centering
 \includegraphics[width=0.5\textwidth, keepaspectratio]{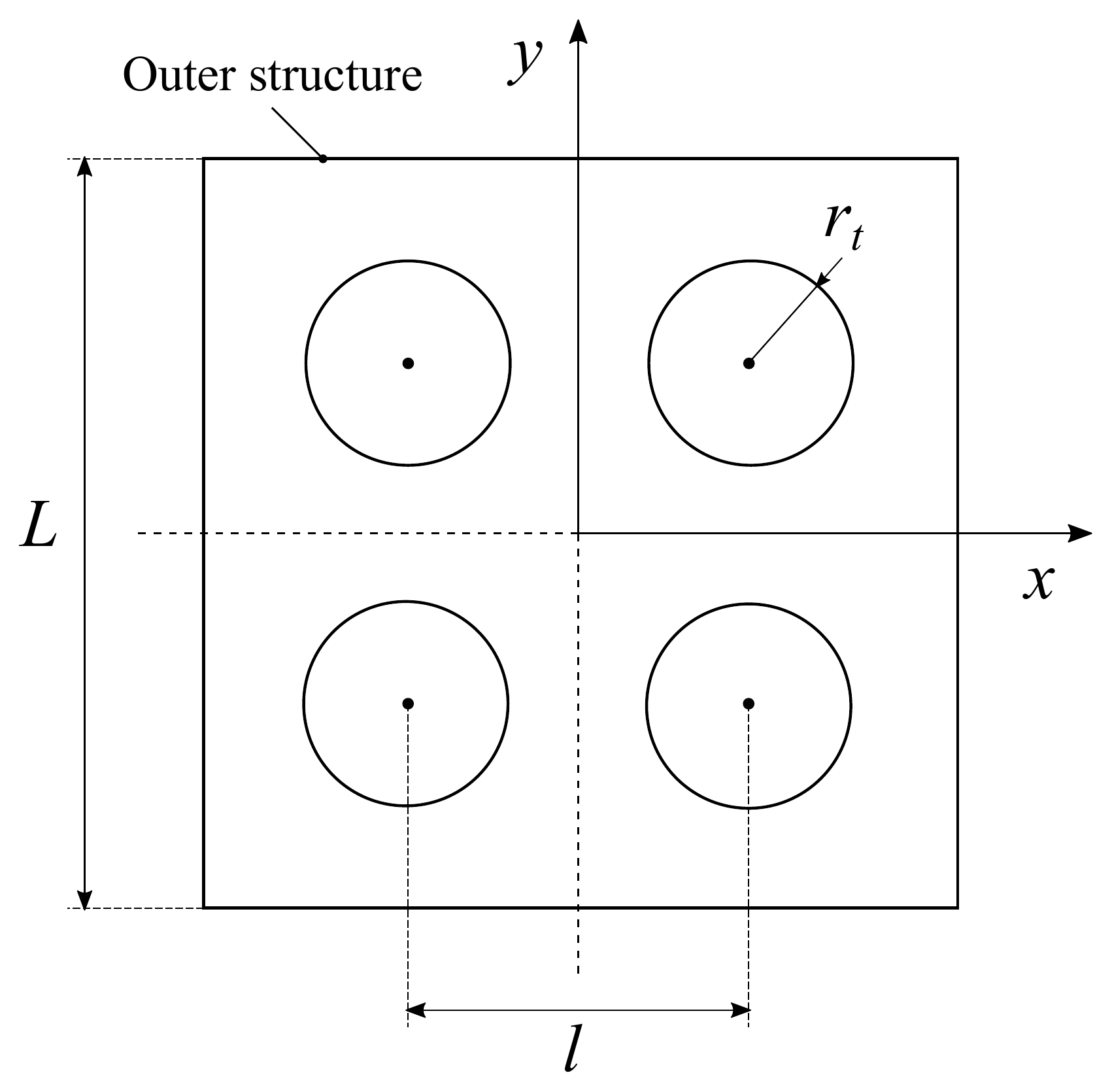}
 	\caption{Example of a tank configuration with four tanks equally spaced with respect to the centre of mass.}
 	\label{fig:tank_pos}
 \end{figure}

For this simplified test case, the positions of the tanks were not optimized and they have been evenly distributed around the barycentre of the spacecraft structure (\cref{fig:tank_pos}).
 
As specified in \cref{subsubsec:tank_constraints}, the tanks need to satisfy strength requirements. Consequently, parameters such as the storage pressure, the safety factor, and the filling factor must be provided. For this specific test case, the storage pressure of the tank is 4 \si{\mega\pascal} \cite{Wertz1999}, the safety factor of the tanks is 1.5 \cite{Wertz1999}, and the filling factor is 1.4 (average value from \cite{AirbusSafranLaunchersGmbH2003}).
\bigbreak
Finally, for the optimisation procedure, it is necessary to provide the mutation and crossover probabilities for the \gls{nsga2} algorithm, as well as the number of individuals in the population and the number of generations used. The crossover and mutation probability are 0.95 and 0.01 respectively. The population size is 80 individuals and the number of generations is 60.

\subsection{Results and discussion}
\label{subsec:results1}
\begin{figure}[hbt!]
	\centering
	\begin{subfigure}[b]{0.48\textwidth}
		\centering
		\includegraphics[height=5.4cm]{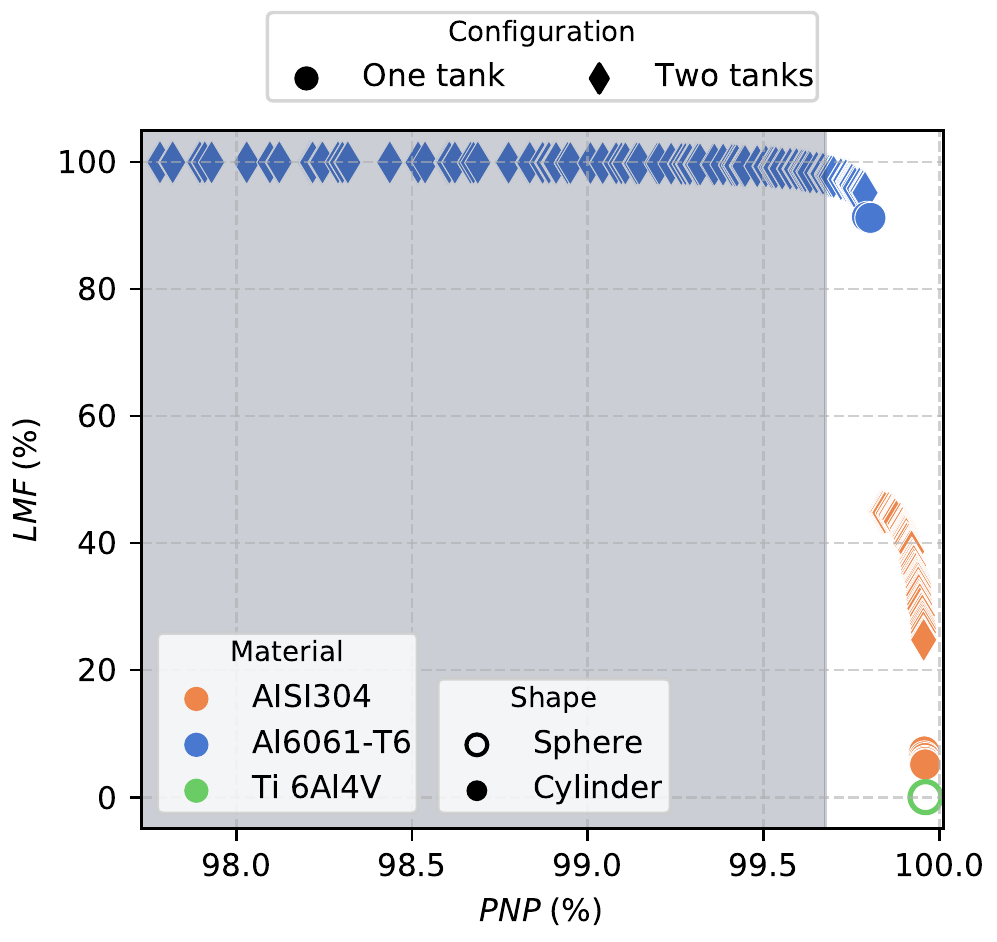}
		\caption{Unconstrained}
		\label{fig:pareto_yr10_nt2_u}
	\end{subfigure}
	\quad
	\begin{subfigure}[b]{0.48\textwidth}
		\centering
		\includegraphics[height=5.4cm]{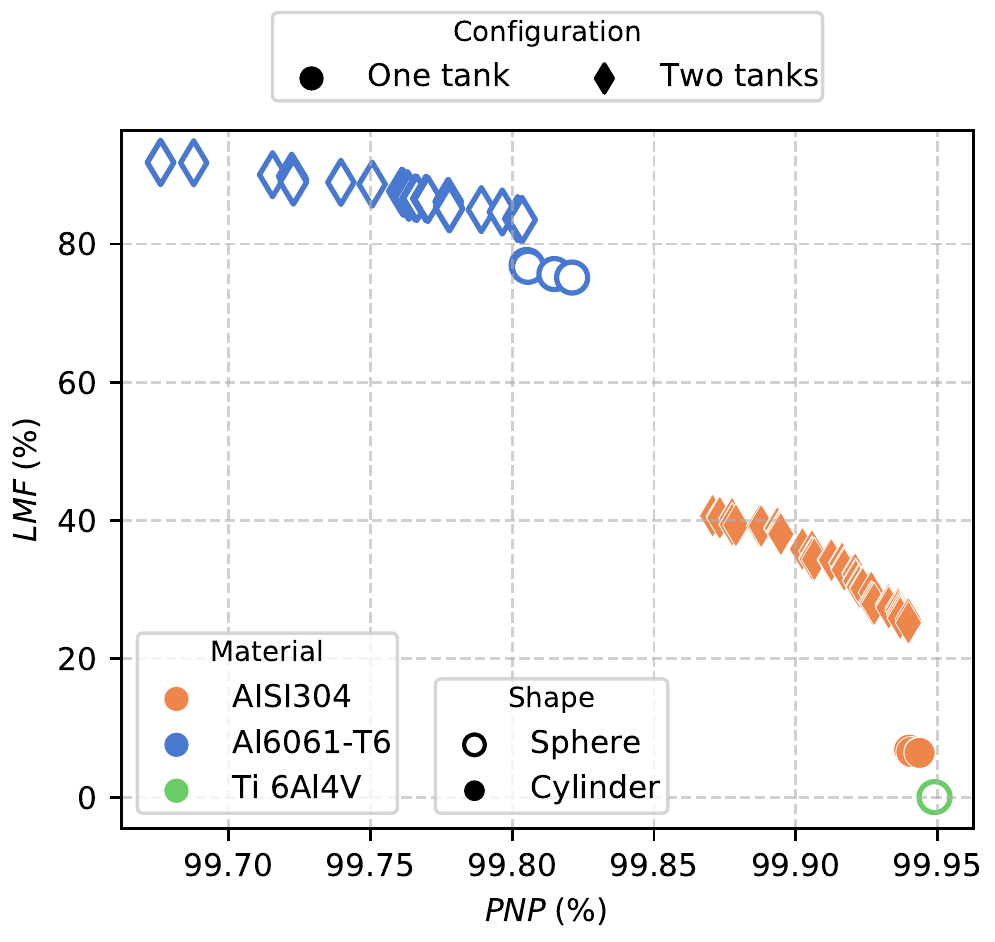}
		\caption{Constrained}
		\label{fig:pareto_yr10_nt2_c}
	\end{subfigure}
	\caption{Pareto front for a 2000 kg spacecraft, 10 years mission lifetime, and 2 maximum tanks.}
	\label{fig:pareto_yr10_nt2}
\end{figure}

\begin{figure}[hbt!]
	\centering
	\begin{subfigure}[b]{0.48\textwidth}
		\centering
		\includegraphics[height=5.4cm]{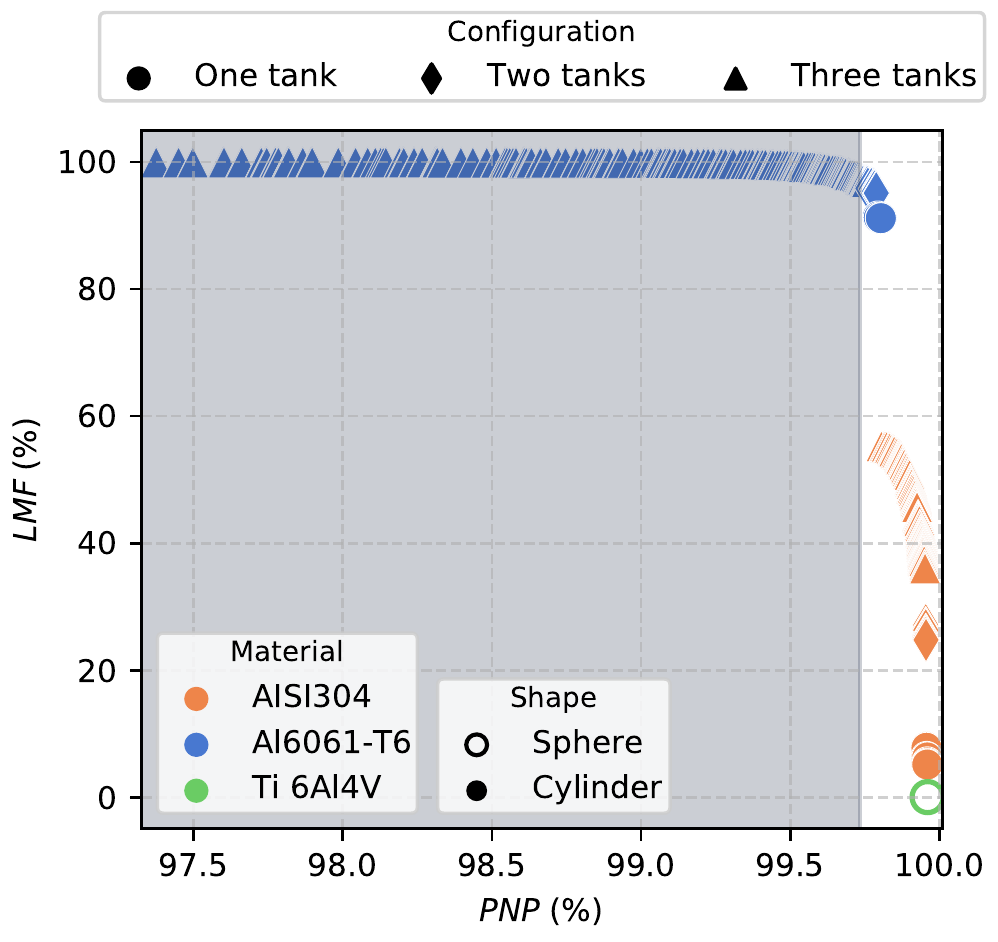}
		\caption{Unconstrained}
		\label{fig:pareto_yr10_nt3_u}
	\end{subfigure}
	\quad
	\begin{subfigure}[b]{0.48\textwidth}
		\centering
		\includegraphics[height=5.4cm]{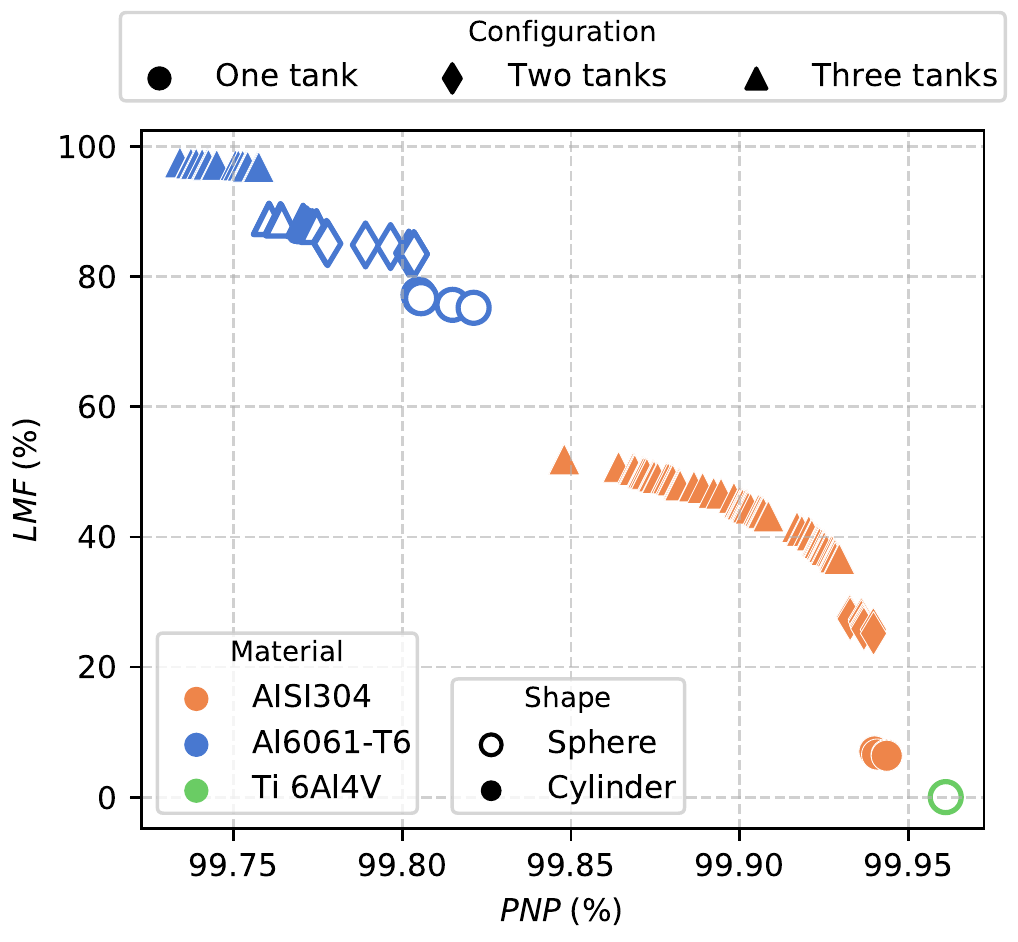}
		\caption{Constrained}
		\label{fig:pareto_yr10_nt3_c}
	\end{subfigure}
	\caption{Pareto front for a 2000 kg spacecraft, 10 years mission lifetime, and 3 maximum tanks.}
	\label{fig:pareto_yr10_nt3}
\end{figure}

The results presented in the following section aim at comparing the output of the optimisation with and without the implementation of the constraints. As previously mentioned, to do so we have used a test case already presented in \cite{Trisolini2018_AESCTE}. The comparison is obtained by showing the Pareto fronts obtained without constraints and comparing them with the corresponding constrained result. For the test case in exam, we compare the Pareto fronts of a 2000 kg spacecraft with a lifetime of 10 years and a maximum allowed number of tanks of 2 and 3 vessels. \cref{,fig:pareto_yr10_nt2,fig:pareto_yr10_nt3} shows the difference between the unconstrained and constrained optimisation results for the mission scenarios. In each Pareto front, the geometries of the markers represent the tank configuration (number of vessels), the colour the material (Aluminium alloy, Stainless-steel, Titanium alloy), and the style of the markers (filled or not) the shape of the tanks (sphere or cylinder). The shaded areas in the unconstrained plots (\cref{fig:pareto_yr10_nt2_u,fig:pareto_yr10_nt3_u}) shows the survivability range of the constrained result with respect to the unconstrained one.

Differences and similarities can be observed comparing the unconstrained and constrained solutions. The first difference is the reduction in the survivability range due to the application of the constraints. In fact, a large part of the high demisability solutions became unfeasible because their thickness is not sufficient to sustain the prescribed operating pressure. The survivability range obtained by the constrained solution may seem narrow; however, values of the \gls{pnp} index for typical missions are in the range 95\%-98\% (obtained as the sum of the contribution of every component in the spacecraft). Consequently, even small variations at component level can be significant, especially when these components are critical to the mission success. For example, the overall probability of no-penetration for the MetOp SVM satellite evaluated with the software SHIELD, is 97.26\%, and the tank assembly (four tanks in total) probability of no-penetration is 99.78\% \cite{Putzar2006}, which is comparable with the value obtained in this work.
The remaining part of the Pareto front obtained with the unconstrained optimisation instead closely resembles the constrained one, in both examples. In particular, the high survivability solutions shows the same behaviour: from left to right, we have a transition from stainless steel solutions with higher number of vessels (more demisable given the smaller sizes) to lower number of vessels, to end with a titanium alloy solutions. Titanium solutions are always dominated in terms of the demisability, given their high temperature resistance. However, among the low demisability solutions, they are more resistant to debris impacts than their stainless-steel equivalents.
Another notable difference can be observed in the feasible high demisability solutions. \cref{fig:pareto_yr10_nt2} shows that the constrained solution only has spherical aluminium tanks. In fact, by limiting the maximum number of vessels to two, the cylindrical aluminium solutions are not feasible anymore and are substituted by spherical ones. A consequence of this is the reduction of the maximum demisability, because spherical tanks are less demisable than cylindrical ones with the same capacity \cite{Trisolini2018_Acta}. As seen in \cref{fig:pareto_yr10_nt2_c}, a two-tank solution does not guarantee full demisability; therefore, in this case, the tanks that present a better \gls{lmf} index, will have a worst performance in terms of casualty area because of the larger cross-section of the two-tank configuration with respect to the single vessel configuration. \cref{fig:pareto_yr10_nt3}, having a maximum number of tanks equal to three, instead shows some cylindrical aluminium solutions. These solutions guarantee the maximum demisability, while still being feasible. Moving from left to right in the Pareto front, we can then observe few three-tanks spherical solutions, which provides lower demisability but higher survivability than the cylindrical counterparts. These solutions are not present in the unconstrained case as they are \emph{replaces} by cylindrical solutions that are now unfeasible. The remaining solutions are two-tanks and one-tank spherical configurations that were also present in \cref{fig:pareto_yr10_nt2}.

Overall, \cref{fig:pareto_yr10_nt2,fig:pareto_yr10_nt3} show how the application of the component specific constraints can change the shape of the Pareto front and the characteristics of the obtained solutions, thus generating readily available, mission compliant, preliminary configurations.


\section{Test case: Medium-to-Large LEO spacecraft}
\label{sec:test2}
This second test case is dedicated to the application of the multi-objective optimisation procedure to a general spacecraft configuration. In the test case of \cref{subsec:results1}, only a single subsystem has been analysed; instead, in this case, multiple components relative to different subsystems have been considered. The resulting optimisation problem becomes more complex as a consequence of a larger optimisation space and the introduction of mission and component specific constraints to be satisfied.

\subsection{Spacecraft configuration}
\label{subsec:config2}
\begin{figure}[htb!]
	\centering
	\includegraphics[width=0.65\textwidth, keepaspectratio]{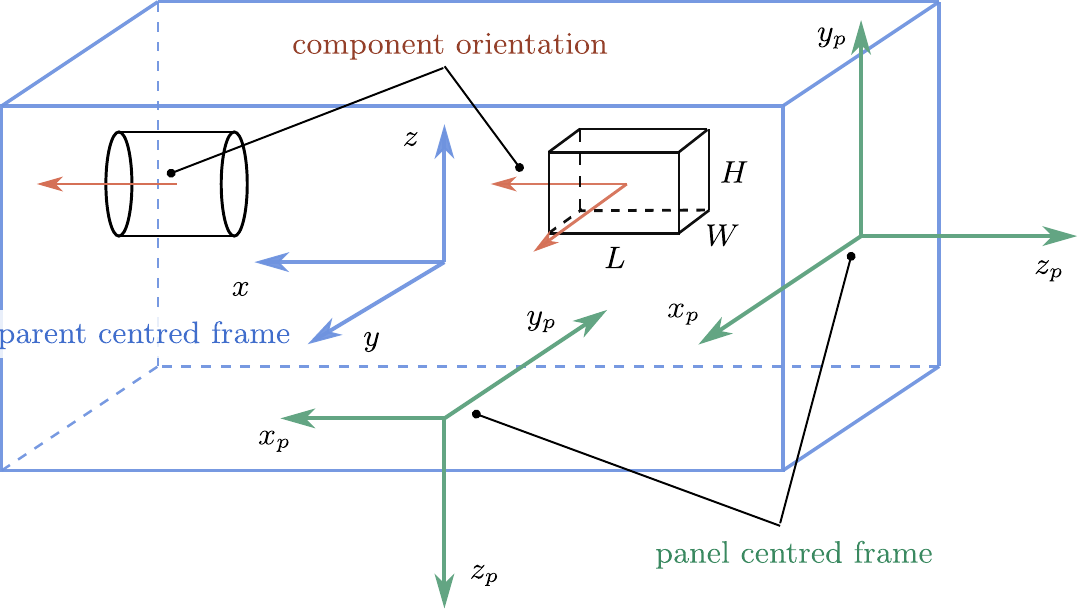}
	\caption{Reference frames used in the definition of the position and orientation of the components.}
	\label{fig:scConfigDef}
\end{figure}

As introduced in \cref{sec:opt_framework}, the optimisation relies on the definition of an baseline configuration, where all the components included in the spacecraft design are specified. During the initialization of the optimisation is possible to specify which components in this configuration will be optimised and, for each component, which variables are going to be optimized. For the variables that will be optimised the provided values serve as place-holders that will be substituted during the generation of the population used in the optimisation. For the test case in exam, \cref{tab:config_parent,tab:config_panels,tab:config_comp} summarise the baseline configurations, describing the parent spacecraft, the external panels and the internal components respectively. In the tables, the symbols and abbreviations have the following meaning: $ID$ is an integer identifier different for each component; \textit{Name} is the name of each object; $ID_p$ is the id of the parent object to which a component is contained or connected; \textit{Shape} defines the geometric shape of the object among Sphere, Box, Flat Plate and Cylinder; \textit{Mat.} is the material of the object; $m$ the thermal mass; $l$ the length of the object that is the dimension along the $x$-axis in the parent centred reference frame (\cref{fig:scConfigDef}); $w$ is the width of the object that is the dimension along the $y$-axis of the parent centred frame; $h$ is the height of the object (along the $z$-axis); $t_s$ is the thickness of the object (if $m$ is provided this is automatically computed); $T_{\rm in}$ is the initial temperature of the object; $m_{\rm aero}$ is the aerodynamic mass; $n$ is the number of objects of the same type.

The parent structure refers to a medium-sized \gls{leo} satellite with an overall mass of 3000 kg, dimensions of 3.4 m x 2.5 m x 2.5 m, and a single 10 m x 2 m solar (\cref{tab:config_parent}). For the main spacecraft structure, only the aerodynamic mass, $m_{aero}$, is provided. It is different from the thermal mass, $m$, which is the mass actually involved in the ablation process. The distinction between the two is crucial in modelling the demise of contained components. In fact, the mass of the internal components affects the aerodynamics but it does not take part in the ablation process. For the parent object, the thermal mass is not provided as it is not taken into account in the demise process. 

\cref{tab:config_panels} shows the configuration definition for the external panels. Separate properties can be defined for the RAM, Trail, Space, Earth, Left, and Right panels defining the external structure of the spacecraft. The size of the panels is provided through the length, width, and thickness. Single-walled panels (FP) and \gls{hcsp} can be defined; this latter case, requires two additional parameters, the thickness of the honeycomb structure ($s_{HC}$) and its areal density ($A\!D_{HC}$).

Finally, \cref{tab:config_comp} summarises the baseline configuration of the internal components. The definition is equivalent to the ones of \cref{tab:config_parent,tab:config_panels}, with few notable differences. For example, component 12, which represents battery cells, is included inside component 11. This is specified through the \emph{parent} variable. The same variable can also be used for identifying the attachment of a component to an external panel; this is the case of component 15, which is attached to the Earth facing panel (ID number 4) as to model an Earth observing payload. Two additional variables are also present: \textit{Pos.}, which is is the position of the component and $\sphericalangle$, which refers to the orientation of the object. As mentioned in \cref{subsec:individuals}, the orientation is defined only with respect to the main axes. For cylindrical objects the orientation is specified with the direction of its axis, while for box-shaped objects two directions are required that are the direction of the length and width sides. The position of the components can be \emph{free-floating} or attached to the external panels. In the first case, the position is defined with respect to the parent spacecraft reference frame, while in the second case with respect to a panel centred frame (\cref{fig:scConfigDef}). 

\begin{landscape}

\begin{table}[hbt!]
\caption{\label{tab:config_parent} Definition of the parent object and the solar panels for the test case configuration.}
\centering
\begin{tabular}{>{\rowmac}l>{\rowmac}l>{\rowmac}c>{\rowmac}c>{\rowmac}c>{\rowmac}c>{\rowmac}c>{\rowmac}c>{\rowmac}c>{\rowmac}c>{\rowmac}c>{\rowmac}c>{\rowmac}c>{\rowmac}c>{\rowmac}c<{\clearrow}}
\hline
\setrow{\bfseries}ID & Name & $\mathbf{ID_{p}}$ & Shape & Mat. & \boldmath{$m$} & \boldmath{$l$} & \boldmath{$r$} & \boldmath{$w$} & \boldmath{$h$} & \boldmath{$t_s$} & \boldmath{$T_{in}$} & \boldmath{$m_{aero}$} & \boldmath{$n$} & \\
& & & & & (\si{\kilo\gram}) & (\si{\meter}) & (\si{\meter}) & (\si{\meter}) & (\si{\meter}) & (\si{\milli\meter}) & (\si{\kelvin}) & (\si{\kilo\gram}) & & \\ \hline
0 & Parent & n/a & Box & Al-6061-T6 & n/a & 3.4 & n/a & 2.5 & 2.5 & 0.002 & 300 & 3000 & n/a & \\
1 & Solar Panel & 0 & Flat Plate & Al-6061-T6 & 82 & 10 & n/a & 2 & n/a & n/a & 300 & 0 & 1 & \\
\hline
\end{tabular}
\end{table}

\begin{table}[hbt!]
\caption{\label{tab:config_panels} Definition of the external panels for the test case configuration.}
\centering
\begin{tabular}{>{\rowmac}l>{\rowmac}l>{\rowmac}c>{\rowmac}c>{\rowmac}c>{\rowmac}c>{\rowmac}c>{\rowmac}c>{\rowmac}c>{\rowmac}c>{\rowmac}c>{\rowmac}c>{\rowmac}c>{\rowmac}c<{\clearrow}}
\hline
\setrow{\bfseries}ID & Name & $\mathbf{ID_{p}}$ & Shape & Mat. & \boldmath{$l$} & \boldmath{$w$} & \boldmath{$t_s$} & \boldmath{$T_{in}$} & \boldmath{$m_{aero}$} & $\boldsymbol{\sphericalangle}$ & \boldmath{$s_{HC}$} & \boldmath{$A\!D_{HC}$} & \\
& & & & & (\si{\meter}) & (\si{\meter}) & (\si{\milli\meter}) & (\si{\kelvin}) & (\si{\kilo\gram}) & (-) & (\si{\milli\meter}) & (\si{\kilo\gram} / \si{\meter\squared}) & \\ \hline
2 & RAM panel & 0 & HC-SP & Graphite-epoxy 1 & 2.5 & 2.5 & 0.001 & 300 & n/d & RAM & 50 & 4.0 \\
3 & Trail panel & 0 & FP & Al-6061-T6 & 2.5 & 2.5 & 0.002 & 300 & n/d & Trail & n/a & n/a \\
4 & Earth panel & 0 & FP & Al-6061-T6 & 3.4 & 2.5 & 0.002 & 300 & n/d & Earth & n/a & n/a \\
5 & Space panel & 0 & FP & Al-6061-T6 & 3.4 & 2.5 & 0.002 & 300 & n/d & Space & n/a & n/a \\
6 & Left panel & 0 & FP & Al-6061-T6 & 3.4 & 2.5 & 0.002 & 300 & n/d & Left & n/a & n/a \\
7 & Right panel & 0 & FP & Al-6061-T6 & 3.4 & 2.5 & 0.002 & 300 & n/d & Right & n/a & n/a \\
\hline
\end{tabular}
\end{table}

\begin{table}[hbt!]
\caption{\label{tab:config_comp} Definition of the internal components for the test case configuration.}
\centering
\begin{tabular}{>{\rowmac}l>{\rowmac}l>{\rowmac}c>{\rowmac}c>{\rowmac}c>{\rowmac}c>{\rowmac}c>{\rowmac}c>{\rowmac}c>{\rowmac}c>{\rowmac}c>{\rowmac}c>{\rowmac}c>{\rowmac}c>{\rowmac}c>{\rowmac}c<{\clearrow}}
\hline
\setrow{\bfseries}ID & Name & $\mathbf{ID_{p}}$ & Shape & Mat. & \boldmath{$l$} & \boldmath{$r$} & \boldmath{$w$} & \boldmath{$h$} & \boldmath{$t_s$} & \boldmath{$T_{in}$} & \boldmath{$m_{aero}$} & Pos. & $\boldsymbol{\sphericalangle}$ &\boldmath{$n$} & \\
& & & & & (\si{\meter}) & (\si{\meter}) & (\si{\meter}) & (\si{\meter}) & (\si{\milli\meter}) & (\si{\kelvin}) & (\si{\kilo\gram}) & & (\si{\meter}) & & \\ \hline
9 & RW & 0 & Cyl. & AISI-316 & 0.06 & 0.15 & n/a & n/a & 0.03 & 300 & 0 & n/d & Left & 1 & \\
10 & Tank & 0 & Cyl. & Ti 6Al4V & 0.6 & 0.30, & n/a & n/a & 0.005 & 300 & 0 & (0.6, 0, 0) & RAM & 1 & \\
11 & BattBox1 & 0 & Box & Al-6061-T6 & 0.6 & n/a & 0.4 & 0.4 & 0.003 & 300 & 0 & n/d & (RAM, Left) & 1 & \\
12 & Batt1 & 11 & Cyl. & Al-6061-T6 & 0.2 & 0.05 & n/a & n/a & 0.001 & 300 & 0 & n/a & n/a & 5 & \\
13 & EBox1 & 0 & Box & Al-6061-T6 & 0.6 & n/a & 0.4 & 0.4 & 0.003 & 300 & 0 & n/d & (RAM, Left) & 1 & \\
14 & EBox2 & 0 & Box & Al-6061-T6 & 0.6 & n/a & 0.4 & 0.4 & 0.003 & 300 & 0 & n/d & (RAM, Left) & 1 & \\
15 & Payload1 & 4 & Box & Al-6061-T6 & 1.0 & n/a & 0.6 & 0.6 & 0.003 & 300 & 0 & n/d & (Left, RAM) & 1 & \\
\hline
\end{tabular}
\end{table}

\end{landscape}

\subsection{Mission scenario}
\label{subsec:mission_scenario}

The mission scenario for this second test case coincides with the scenario of \cref{subsec:results1}, therefore we maintain the same operational orbit for the survivability analysis and the same initial conditions for the demisability analysis. Alongside the mission scenario, additional information about the mission design must be provided as it is relevant for the definition of the constraints. \cref{tab:mission_design} summarises the main design parameters selected for the test case in exam.

\begin{table}[hbt!]
	\caption{\label{tab:mission_design} Design parameter for the main mission components.}
	\centering
	\begin{tabular}{llccc}
		\hline
		\textbf{Component} & \textbf{Design parameter} & \textbf{Symbol} & \textbf{Value} & \textbf{Unit} \\
		\hline
		\multirow{5}*{Tank} & Propellant mass 		& $m_f$ 	& 220 	& \si{\kilo\gram} \\
		& Propellant density 	& $\rho_f$	& 1.02 	& \si{\kilo\gram / \meter\cubic} \\
		& Max. pressure			& $p_{\rm max}$		& 4		& \si{\mega\pascal} \\
		& Safety factor 		& $SF$ 		& 1.5	& \\
		& Filling factor		& $K_1$ 	& 1.4 	& \\
		\hline
		\multirow{4}*{Reaction wheel} 	& Angular momentum 		& $H_d$ 	& 85 	& \si{\newton\meter\second} \\
		& Max. angular speed	& $\omega_{\rm max}$ 	& 5000 & rpm \\
		& Safety factor 		& $SF$		& 1.5	& \\
		& Aspect ratio 			& $AR$		& 0.2 	& \\
		\hline
		\multirow{2}*{Battery cells}	& Eclipse time	& $T_e$ & 35.13 & \si{\minute} \\
		& Eclipse power & $W_e$ & 2.1 & \si{\kilo\watt} \\							
		\hline
	\end{tabular}
\end{table}

As in \cref{sec:test1}, the number of tanks is optimised but not the position; consequently, the centre of mass of the assembly has fixed coordinates of (0.6 m, 0 m, 0 m). The values of the required propellant mass, angular momentum, and required power have been obtained through linear interpolation of satellite data gathered from the database of the Union of Concerned Scientists \cite{UnionofConcernedScientists2017} as a function of the satellite mass. For a more details the reader can refer to \cite{Trisolini_Thesis}.

\subsection{optimisation set-up}
\label{subsec:opt_setup2}
Through the optimisation framework it is possible to select the components to be optimised and, for each of them, the relevant optimisation variables, together with their boundaries and options. \cref{tab:opt_variables2} summarises the components optimised for the test case in exam. All internal components have been considered for optimisation. In addition, the RAM panel has been selected. This choice is based considering that the RAM panel is the most exposed to debris impacts. For continuous variables, the upper and lower bounds are displayed, while for integer variables the possible options are indicated. The material was always included in the optimisation as it is one of the most influential parameters for both the demisability and the survivability \cite{Trisolini2018_Acta}. For the case in exam, we used a selected number of materials (Al 6061-T6, AISI-316, Ti 6Al4V, Graphite epoxy), but it is possible to extend the list to any amount. For some of the components, the material is constrained; for example, the RAM panel can be either aluminium or graphite epoxy as these are the types of material used for external panels. Some components, such as reaction wheels and battery boxes have been constrained to be attached to external panels. For example, reaction wheels can be attached to any panels, while the battery box with ID 11 can only be attached to the Space and Right panels. In addition, for some components even the position inside the spacecraft has been limited in order to test the position error handling algorithm of \cref{subsubsec:pos_constraints}. As an example, the payload component (ID 17) can only be attached to the Earth panel and occupy a limited area on it.

\begin{spacing}{1.}
\begin{longtable}[c]{l|c|c|c}
\caption{\label{tab:opt_variables2} Optimisation variables for each component with the lower and upper bounds or the available options.} \\
\hline
\textbf{Comp.} & \textbf{ID} & \textbf{Variable} & \textbf{Bounds / Options} \\
\hline
\multirow{3}*{RAM Panel} & \multirow{3}*{2} & Mat. & Al-6061-T6, Graphite-epoxy 1 \\ & & Shape & \gls{fp}, \gls{hcsp} \\ & & $t_s$ & 0.0004 \si{\meter} - 0.003 \si{\meter} \\
\hline
\multirow{5}*{RW} & \multirow{4}*{9} & Mat. & Al-6061-T6, AISI-316, Ti 6Al4V \\ & & $r$ & 0.05 \si{\meter} - 0.20 \si{\meter} \\ & & $\mathrm{ID_{p}}$ & 2 - 7 \\ & & \multirow{2}*{Pos.} & -1.7 \si{\meter} $\leq x \leq$ 1.7 \si{\meter} \\ & & & -1.25 \si{\meter} $\leq y \leq$ 1.25 \si{\meter} \\
\hline
\multirow{4}*{Tank} & \multirow{4}*{10} & Mat. & Al-6061-T6, AISI-316, Ti 6Al4V \\ & & Shape & Sphere, Cylinder \\ & & $t_s$ & 0.0005 \si{\meter} - 0.003 \si{\meter} \\ & & $n_t$ & 1 - 4 \\
\hline
\multirow{5}*{BattBox1} & \multirow{4}*{11} & Mat. & Al-6061-T6, Ti 6Al4V \\ & & $t_s$ & 0.001\si{\meter} - 0.003\si{\meter} \\ & & $\mathrm{ID_{p}}$ & 5, 7 \\ & & \multirow{2}*{Pos.} & -1.7 \si{\meter} $\leq x \leq$ 1.7 \si{\meter} \\ & & & -1.25 \si{\meter} $\leq y \leq$ 1.25 \si{\meter} \\
\hline
\multirow{2}*{Batt1} & \multirow{2}*{12} & Mat. & Al-6061-T6, AISI-316 \\ & & Catalogue & Table \ref{tab:batt_cell} \\
\hline
\multirow{6}*{EBox1} & \multirow{6}*{13} & Mat. & Al-6061-T6, Ti 6Al4V \\ & & $\mathrm{ID_{p}}$ & 0, 2, 7 \\ & & $t_s$ & 0.001 \si{\meter} - 0.003 \si{\meter} \\ & & \multirow{3}*{Pos.} & -1.7 \si{\meter} $\leq x \leq$ 1.7 \si{\meter} \\ & & & -1.7 \si{\meter} $\leq y \leq$ 1.7 \si{\meter} \\ & & &  -1.25 \si{\meter} $\leq z \leq$ 1.25 \si{\meter} \\
\hline
\multirow{6}*{EBox2} & \multirow{6}*{14} & Mat. & Al-6061-T6, Ti 6Al4V \\ & & $\mathrm{ID_{p}}$ & 0, 2, 5 \\ & & $t_s$ & 0.001 \si{\meter} - 0.003 \si{\meter} \\ & & \multirow{3}*{Pos.} & -1.7 \si{\meter} $\leq x \leq$ 1.7 \si{\meter} \\ & & & -1.7 \si{\meter} $\leq y \leq$ 1.7 \si{\meter} \\ & & &  -1.25 \si{\meter} $\leq z \leq$ 1.25 \si{\meter} \\
\hline
\multirow{3}*{Payload1} & \multirow{3}*{15} & $t_s$ & 0.001 \si{\meter} - 0.003 \si{\meter} \\ & & \multirow{2}*{Pos.} & 0.7 \si{\meter} $\leq x \leq$ 1.7 \si{\meter} \\ & & & -1.25 \si{\meter} $\leq y \leq$ 1.25 \si{\meter} \\
\hline
\end{longtable}
\end{spacing}

Another feature demonstrated with this test case is the possibility to pick components from a catalogue. The test has been performed with battery cells, which could be selected from the catalogue of \cref{tab:batt_cell}. The five different cell types are characterised by different masses, shapes, and dimensions. It is assumed that the cells geometries of \cref{tab:batt_cell} can be used for both Li-ion and Ni-Cd cell types. The difference between the two types of batteries is considered through the material: Li-ion batteries have aluminium alloy casing while Ni-Cd have stainless-steel casings. In addition, for both cell types is assumed a constant energy density and \gls{dod}. For Li-ion cells, the energy density is 140 \si{\watt}\si{\hour}/\si{\kilo\gram} and the \gls{dod} is 0.2, while for Ni-Cd cells, the energy density is 60 \si{\watt}\si{\hour}/\si{\kilo\gram} and the \gls{dod} is 0.6 \citep{EaglePicherTechnologies2016,Wertz1999}. 

\begin{table}[hbt!]
\caption{\label{tab:batt_cell} Catalogue of battery cells used in the optimisation \cite{EaglePicherTechnologies2016}.}
\centering
{\renewcommand{\arraystretch}{1.1}%
\begin{tabular}{lccc}
\hline
\textbf{Battery ID} & \textbf{m (\si{\kilo\gram})} & \textbf{Shape} & \textbf{Dimensions (\si{\milli\meter})} \\ \hline
0 & 0.38 & Box & $l \: 88 \times w \: 55 \times h \: 39$ \\
1 & 1.15 & Cylinder & $l \: 245 \times \diameter \: 54$ \\
2 & 1.82 & Box & $l \: 175 \times w \: 81 \times h \: 65.2$ \\
3 & 2.2 & Box & $l \: 210 \times w \: 110 \times h \: 76$ \\
4 & 4.5 & Box & $l \: 220 \times w \: 170 \times h \: 95$ \\
\hline
\end{tabular}}
\end{table}

Finally, the parameters necessary for the optimisation algorithm must be provided. The crossover and mutation probability are kept unchanged from the previous test case (Section \ref{subsec:opt_var1}) with a respective value of 0.95 and 0.01. The size of the population is 120 individuals and the number of generations is 100.
 
\subsection{Results and discussion}
\label{subsec:results2}
\begin{figure}[htb!]
\centering
\includegraphics[width=0.75\textwidth]{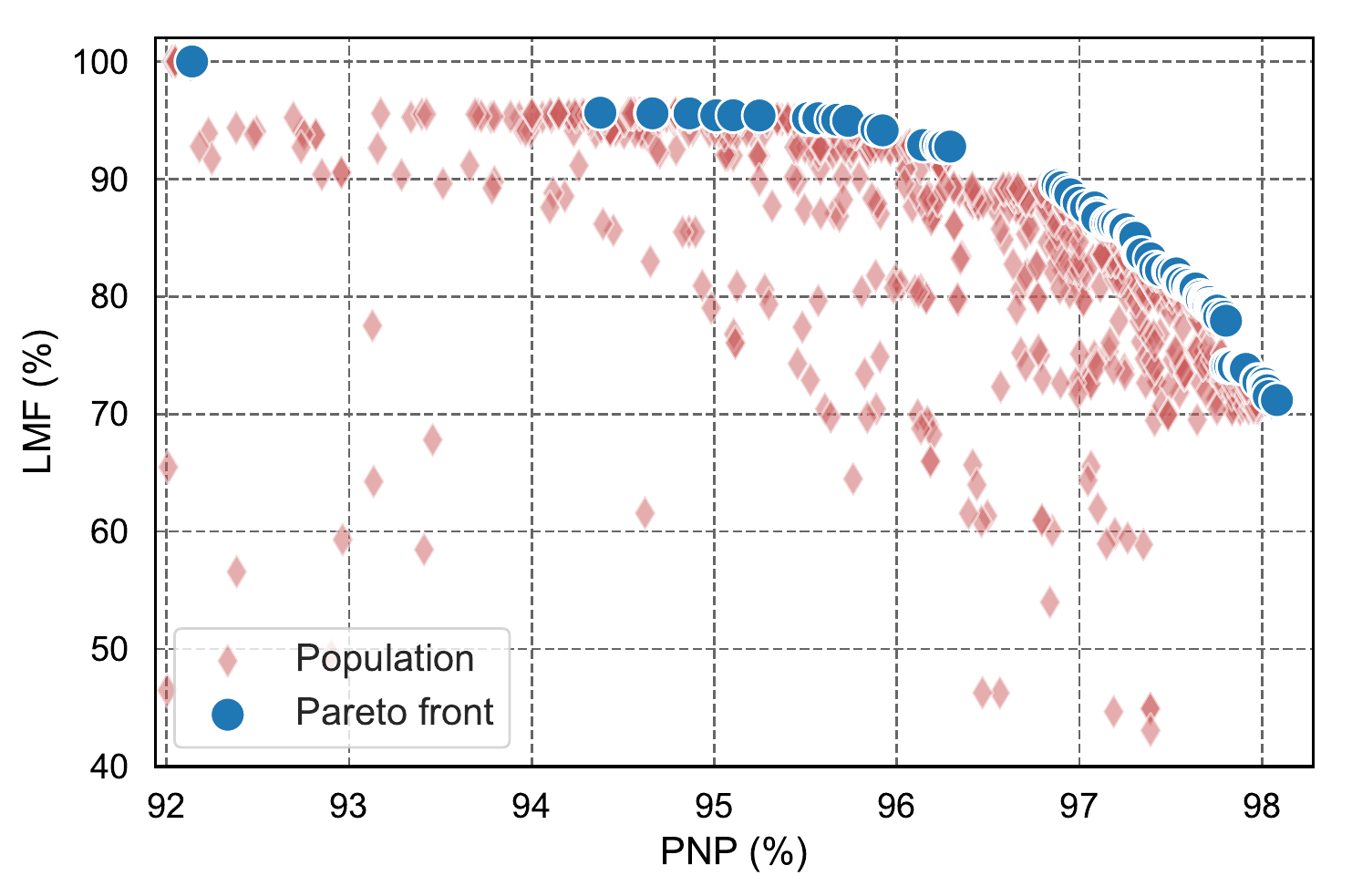}
	\caption{Pareto front for the general spacecraft configuration.}
	\label{fig:pareto_test_2}
\end{figure}

The test cases outlined in the \cref{subsec:config2,subsec:mission_scenario,subsec:opt_setup2}, demonstrates the application of the multi-objective optimisation framework described in \cref{sec:opt_framework}, for a mid-to-large LEO spacecraft. This test case includes many of the features of interest of the framework: the ability to consider different types of components and their respective constraints. The constraints can be both driven by the physical limitations of the components, such as the minimum thickness of a tank vessel, or given by the user, e.g. the necessity of limiting the position of a component to a specific area of the spacecraft. The results obtained from the optimisation are summarised by the Pareto front of \cref{fig:pareto_test_2}.

The first noticeable feature of the Pareto front with respect to the results of \cref{sec:test1} are the different ranges spanned by both the demisability and the survivability indices. The \gls{lmf} index shows a smaller range with respect to the previous test case, with a minimum value of 70\%, indicating that at least 70 percent of the mass of the considered components demises during the re-entry phase. However, given the greater number of components, also the survivability index increases its range. The \gls{pnp} index ranges from about 92\% to slightly above 98\%. This is a considerable difference in the spacecraft ability to survive impacts during the mission lifetime and must be taken into account. A fully demisable configuration as found by the optimiser for this test case would likely need further consideration as the vulnerability of the spacecraft would be too high for a typical mission scenario. 

\begin{landscape}
\begin{table}[hbt!]
\caption{\label{tab:pareto_first} Solution of the optimisation with the best demisability.}
\centering
\begin{tabular}{>{\rowmac}l>{\rowmac}l>{\rowmac}c>{\rowmac}c>{\rowmac}c>{\rowmac}c>{\rowmac}c>{\rowmac}c>{\rowmac}c>{\rowmac}c>{\rowmac}c>{\rowmac}c>{\rowmac}c>{\rowmac}c<{\clearrow}}
\hline
\setrow{\bfseries}ID & Name & $\mathrm{ID_{p}}$ & Shape & Material & \boldmath{$m$} & \boldmath{$l$} & \boldmath{$r$} & \boldmath{$w$} & \boldmath{$h$} & \boldmath{$t_s$} & Position &\boldmath{$n$} & \\
& & & & & (\si{\kilo\gram}) & (\si{\meter}) & (\si{\meter}) & (\si{\meter}) & (\si{\meter}) & (\si{\milli\meter}) & (\si{\meter}) & & \\ \hline
2 & RAM Panel & 0 & \gls{hcsp} & Graphite-epoxy 1 & n/a & 2.5 & n/a & 2.5 & n/a & 0.00245 & n/a & n/a & \\
9a & RW & 5 & Cyl. & Al-6061-T6 & n/a & 0.0528 & 0.132 & n/a & n/a & 0.0264 & (0.155, 0.929) & 1 & \\
9b & RW & 6 & Cyl. & Al-6061-T6 & n/a & 0.0528 & 0.132 & n/a & n/a & 0.0264 & (-1.286, 0.313) & 1 & \\
9c & RW & 2 & Cyl. & Al-6061-T6 & n/a & 0.0528 & 0.132 & n/a & n/a & 0.0264 & (-0.45, -1.012) & 1 & \\
10a & Tank & 0 & Sphere & Al-6061-T6 & n/a & n/a & 0.291 & n/a & n/a & 0.00292 & (0.9, 0, 0) & 1 & \\
10b & Tank & 0 & Sphere & Al-6061-T6 & n/a & n/a & 0.291 & n/a & n/a & 0.00292 & (0.297, 0, 0.350) & 1 & \\
10c & Tank & 0 & Sphere & Al-6061-T6 & n/a & n/a & 0.291 & n/a & n/a & 0.00292 & (0.297, 0, -0.350) & 1 & \\
11 & BattBox1 & 7 & Box & Ti 6Al4V & n/a & 0.798 & n/a & 0.231 & 0.2205 & 0.0021 & (1.289, 0.103) & 1 & \\
12 & Batt1 & 11 & Box & Al-6061-T6 & 2.2 & 0.21 & n/a & 0.11 & 0.076 & n/a & n/a & 19 & \\
13 & EBox1 & 2 & Box & Al-6061-T6 & n/a & 0.6 & n/a & 0.4 & 0.4 & 0.0025 & (0.915, -0.417) & 1 & \\
14 & EBox2 & 0 & Box & Al-6061-T6 & n/a & 0.6 & n/a & 0.4 & 0.4 & 0.0027 & (1.354, 0.561, 0) & 1 & \\
15 & Payload1 & 4 & Box & Al-6061-T6 & n/a & 1.0 & n/a & 0.6 & 0.6 & 0.00296 & (1.293, -0.693) & 1 & \\
\hline
\end{tabular}
\end{table}

\begin{table}[hbt!]
\caption{\label{tab:pareto_last} Solution of the optimisation with the best survivability.}
\centering
\begin{tabular}{>{\rowmac}l>{\rowmac}l>{\rowmac}c>{\rowmac}c>{\rowmac}c>{\rowmac}c>{\rowmac}c>{\rowmac}c>{\rowmac}c>{\rowmac}c>{\rowmac}c>{\rowmac}c>{\rowmac}c>{\rowmac}c<{\clearrow}}
\hline
\setrow{\bfseries}ID & Name & $\mathrm{ID_{p}}$ & Shape & Material & \boldmath{$m$} & \boldmath{$l$} & \boldmath{$r$} & \boldmath{$w$} & \boldmath{$h$} & \boldmath{$t_s$} & Position &\boldmath{$n$} & \\
& & & & & (\si{\kilo\gram}) & (\si{\meter}) & (\si{\meter}) & (\si{\meter}) & (\si{\meter}) & (\si{\milli\meter}) & & (\si{\meter}) & \\ \hline
2 & RAM Panel & 0 & \gls{hcsp} & Graphite-epoxy 1 & n/a & 2.5 & n/a & 2.5 & n/a & 0.00244 & n/a & \\
9a & RW & 5 & Cyl. & Al-6061-T6 & n/a & 0.0528 & 0.132 & n/a & n/a & 0.0264 & (-1.1, 1.02) & 1 & \\
9b & RW & 6 & Cyl. & Al-6061-T6 & n/a & 0.0528 & 0.132 & n/a & n/a & 0.0264 & (-1.18, 0.313) & 1 & \\
9c & RW & 2 & Cyl. & Al-6061-T6 & n/a & 0.0528 & 0.132 & n/a & n/a & 0.0264 & (0.46, -1.04) & 1 & \\
10 & Tank & 0 & Sphere & Ti 6Al4V & n/a & n/a & 0.42 & n/a & n/a & 0.0029 & (0.6, 0, 0) & 1 & \\
11 & BattBox1 & 7 & Box & Ti 6Al4V & n/a & 0.53 & n/a & 0.231 & 0.092 & 0.0021 & (1.289, 0.12) & 1 & \\
12 & Batt1 & 11 & Box & Al-6061-T6 & 0.38 & 0.088 & n/a & 0.055 & 0.039 & n/a & n/a & 110 & \\
13 & EBox1 & 2 & Box & Ti 6Al4V & n/a & 0.6 & n/a & 0.4 & 0.4 & 0.0025 & (0.596, -0.417) & 1 & \\
14 & EBox2 & 0 & Box & Ti 6Al4V & n/a & 0.6 & n/a & 0.4 & 0.4 & 0.0028 & (1.354, 1.354, 0.637) & 1 & \\
15 & Payload1 & 4 & Box & Al-6061-T6 & n/a & 1.0 & n/a & 0.6 & 0.6 & 0.00296 & (1.314, -0.693) & 1 & \\
\hline
\end{tabular}
\end{table}

\end{landscape}

\cref{tab:pareto_first} and \cref{tab:pareto_last} show two configurations extracted from the Pareto front of \cref{fig:pareto_test_2}. Specifically, they refer to the first and last solutions, i.e. the solution with the best demisability and the best survivability respectively. Some aspects of the optimisation can be observed from the two tables and from examining the intermediate solutions. The first aspect is that the shape and behaviour of the Pareto front is mainly dominated by the more massive components for which the material was an optimisation variable that are the tanks, and electronic boxes. For the most demisable solution (\cref{tab:pareto_first}), we have a configuration with three spherical aluminium tanks, which guarantee a fully demisable solution, and aluminium electronic boxes. On the other hand, the solution with the best survivability (\cref{tab:pareto_last}), shows a configuration with only one large spherical titanium tank, and two electronic boxes, also in titanium. The maximum survivability is achieved, as expected, with thickness close to the maximum thickness possible for both the tanks and the electronic boxes.
Another characteristics that can be observed from the two configuration is that both the most demisable and the most resilient configuration have aluminium reaction wheels. This can indicate that the demisable option for the reaction wheels for this spacecraft configuration could only be achieved switching to aluminium. Moreover, this change in material, does not significantly affect the overall survivability of the spacecraft, when measured with the \gls{pnp} index. This is most probably related to the limited dimensions of the reaction wheels: despite using aluminium alloy increases the dimension of the reaction wheels to obtain the design angular momentum, such an increase in dimensions does not significantly reduce the survivability. It has to be noted here that this is a \emph{what if} scenario as, currently, no aluminium reaction wheels have been used on board of satellites. Nonetheless, with respect to their stainless-steel and titanium alloy counterparts they are significantly more demisable and provide a feasible solution in terms of provided angular momentum and strength requirements. The payload component is close to the maximum thickness in both the configurations, thus indicating that it always demises even when the maximum thickness is taken, which on the other hand ensures the maximum possible survivability. Another aspect that can be observed is that almost all components, in both configurations, are clustered around the tank assembly so that the vulnerability of the tank assembly is reduced by their shielding contribution.
In addition, the EBox2 component has in all the solutions in the Pareto front a thickness greater than the EBox1 (when they are of the same material). The two boxes are identical in shape; however, the EBox1 is always attached to the RAM panel, which is protecting it in a grater fashion. Despite the RAM panel collects the largest amount of particles, it is also the most resistant to impacts as it is the only honeycomb sandwich panel in the configuration and the thickness of each plate is comparable to the one of the other single layer panels (2.45 \si{\milli\meter} to 3 \si{\milli\meter} respectively).
\bigbreak


\section{Conclusions and Discussion}
\label{sec:conclusions}

The paper has presented the development of an optimisation framework allowing the trade-off analysis between the demisability and the survivability of preliminary design solutions. The framework has been applied to two relevant test cases: the optimisation of a spacecraft subsystem that is the tank assembly in \cref{sec:test1} and the optimisation of a mid-to-large LEO spacecraft in \cref{sec:test2}. The presented results are based on previous work (\cite{Trisolini2018_AESCTE}) and they expand on it. While in the previous work the optimisation was unconstrained and several unfeasible solutions were found, the included constraints implementation has improved the reliability and usability of the framework as only feasible solutions are returned. The second test case showed the application to a more realistic spacecraft configuration, which included several subsystems (tank assembly, reaction wheel, batteries) and other generic components, with flexibility in the definitions of mission and component related constraints. The framework allows the user to have control on several aspects of the preliminary configuration and so that a feasible and usable solution is found. The framework found a range of solutions as shown in \cref{fig:pareto_test_2}, which span several levels of demisability and survivability from which preliminary configurations can be chosen. However, as mentioned in \cref{sec:test2}, the Pareto front is dominated by the more massive components and some of them may be \emph{overlooked} and their contribution neglected inside the optimisation routine. For example, the reaction wheels have aluminium alloy material in both the most demisable (\cref{tab:pareto_first}) and most survivable (\cref{tab:pareto_last}) solution. This is connected also to the selection of the fitness functions for the optimisation as probably the simple implementation of the \gls{pnp} index cannot adequately distinguish between these solutions. This could be improved, for example, by introducing a criticality coefficient that would assign higher weights to the most critical components, thus increasing the possibility to evaluate their contribution even if they are less massive. Another aspect to be considered is the selection of the optimisation algorithm. At the moment a simple implementation of the \gls{nsga2} algorithm has been implemented, but other heuristic search algorithms, such as differential evolution or particle swarm to name a few, should be investigated as they may have better performances.

\section*{Acknowledgements}
This work was funded by EPSRC DTP/CDT through the grant number EP/K503150/1. It was also supported by the European Research Council (ERC) under the European Union’s Horizon 2020 research and innovation programme (grant agreement No 679086 - COMPASS)

\bibliographystyle{model6-num-names}
\bibliography{mybib}

\appendix

\begin{landscape}
\section{Material database}
\label{sec:mat_database}
\cref{tab:mat_database} shows an extract of the materials available in the code, for both the demisability and the survivability models. The majority of the data is from the Debris Assessment Software 2.0 \cite{NASA2015_DAS,Owens2014} database. When material properties such as the emissivity were not available in the DAS database, they were retrieved from the MatWeb database \cite{MatWebLLC2015} or from other publications \cite{Beck2015,Beck2015b,Lips2005,Lips2005a,Lips2005b,Lips2015}

\begin{table}[hbt!]
\caption{\label{tab:mat_database} Material database of the demisability and survivability models.}
\centering
\begin{tabular}{lccccccccc}
\hline
\textbf{Material} & \bm{$\rho_m$} & \bm{$HB$} & \bm{$T_m$} & \bm{$h_f$} & \bm{$C_m$} & \bm{$\epsilon$} & \bm{$C$} & \bm{$\sigma_y$} & \\ 
 & (\si{\kilo\gram\per\meter\square}) &  & (\si{\kelvin}) & (\si{\joule\per\kilo\gram}) & (\si{\joule\per\kilo\gram\per\kelvin}) &  & (\si{\meter\per\second}) & (\si{\mega\pascal}) & \\ 
\hline
Al-6061-T6 & 2713 & 95 & 867 & 386116 & 896 & 0.141 & 5100 & 276 & \\
Al-7075-T6 & 2787 & 150 & 830 & 376788 & 1012.35 & 0.141 & 5040 & 450 & \\
Titanium 6Al4V & 4437 & 334 & 1943 & 393559 & 805.2 & 0.302 & 4987 & 880 & \\
AISI304 & 7900 & 123 & 1700 & 286098 & 545.1 & 0. 35 & 5790 & 215 & \\
AISI316 & 8026.85 & 149 & 1644 & 286098 & 460.6 & 0.35 & 5790 & 250 & \\
Inconel-601 & 8057.29 & n/a & 1659 & 311664 & 632.9 & 0.122 & 5700 & 450 & \\
Graphite-epoxy 1 & 1570 & n/a & 700 & 1.60E+07 & 1100 & 0.86 & n/a & 498.5 & \\
Graphite-epoxy 2 & 1550.5 & n/a & 700 & 236 & 879 & 0.9 & n/a & 498.5 & \\
\hline
\end{tabular}
\end{table}

For this work, all the material properties are assumed temperature independent. The quantity $\rho_m$ represents the density of the considered material. The Brinell hardness ($HB$) measures the indentation hardness of materials through the scale of penetration of an indenter, loaded on a material test-piece. This parameter is only used in the vulnerability analysis. For this work, the ambient temperature HB hardness has been considered as the BLEs correlations have been obtained with ambient temperature on-ground testing. The melting temperature ($T_m$) and the heat of fusion ($h_f$) are used in the demisability analysis. In the code, the melting temperature is the temperature after which the material starts to demise. After $T_m$ has been reached, the temperature is kept constant and the object starts to demise, losing mass at a rate that is proportional to the heat flux and the heat of fusion. The heat of fusion ($h_f$) represents the amount of energy required to change the state of a substance from solid to liquid at constant pressure. The specific heat capacity ($C_m$) is again only used in the demisability analysis. 
\end{landscape}

It is the amount of energy required to raise the temperature of a substance per unit of mass. The specific heat capacity is a temperature dependent material. As we are using a database with constant quantities, the adopted value is the mean specific heat capacity between the reference and melting temperatures. These ranges are directly provided in the DAS 2.0 database. The emissivity ($\epsilon$) of a material is its effectiveness in emitting energy as thermal radiation. This is an important parameter for the assessment of the demisability of an object. In fact, it regulates the amount of heat that the surface of an object emits as a function of its temperature. During re-entry, the surface of an object reaches high temperature, making the knowledge of the emissivity important for the correct analysis of the heat balance on the re-entering object. In general, the emissivity depends on the temperature and the state of the surface of an object. Given the harsh thermochemical environment of atmospheric re-entries, such state can be altered, for example, by the formation of oxides, which can strongly affect the emissivity of a material, usually increasing it. In the present study, the emissivity is considered constant and relative to a surface that has not been altered by thermochemical reactions. In a future development of the tool we will consider these aspects and include temperature dependant relations for both the emissivity, given the recent efforts in the characterisation of common spaceflight materials for demisability application \cite{pagan2015experimental,pagan2016total}. The parameter $C$ instead represents the speed of sound in the specific material considered. .The yield strength ($\sigma_y$) is defined as the stress at which a material begins to deform plastically. This parameter is only used in the vulnerability analysis for the computation of the critical impact diameter using the BLEs. Similarly, to the Brinell hardness, as the BLEs correlations have been obtained through ground-testing at ambient temperature, the ambient temperature value of the yield strength has been considered.

\end{document}